\documentclass[]{spie}

\usepackage{amsmath,amsfonts,amssymb}
\usepackage{graphicx}
\usepackage[colorlinks=true, allcolors=blue]{hyperref}
\usepackage{aas_macros}

\title{Searching for Gravitational Wave Events Counterparts in X-Rays by Tilting the Multiple Shells of Wolter-I Optics}

\author[a,b,*]{John Rankin}
\author[a]{Daniele Spiga}
\author[a]{Paolo Conconi}
\author[a]{Sergio Campana}
\author[a]{Giovanni Pareschi}
\author[a]{Stefano Basso}
\author[a]{Marta Maria Civitani}
\author[a]{Vincenzo Cotroneo}

\affil[a]{Istituto Nazionale di Astrofisica / Osservatorio Astronomico di Brera (INAF/OAB), Via E. Bianchi, 46, 23807, Merate (LC), Italy}
\affil[b]{Dipartimento di Fisica e Scienze della Terra, Universit\`a degli Studi di Ferrara, Via Saragat 1, 44122, Ferrara, Italy}

\authorinfo{Further author information: (Send correspondence to J.R)\\J.R.: E-mail: john.rankin@unife.it,}

\begin{document} 
\maketitle

\begin{abstract}
Current X-ray telescopes have a combination of field of view and angular resolution that makes it difficult to quickly localize the electromagnetic counterparts of gravitational wave events — which will become increasingly important when the LIGO and Virgo experiment will work at full sensitivity or when the Einstein Telescope and the Cosmic Explorer observatories will come online in the next decades.
   We investigate what X-ray optical designs could provide --- in a small mission --- a field of view wide enough to observe quickly the region of the sky constrained by gravitational wave detectors, and at the same time an angular resolution sufficiently good to localize these sources.
   We study, using ray tracing simulations, how we can optimize a Wolter-I telescope for this purpose. We find that, by appropriately tilting the shells of a Wolter-I X-ray telescope in different directions, the field of view increases, at the cost of a reduction of the on-axis performance.  
  We draw a possible design made of two identical modules pointing at adjacent regions of the sky. 
\end{abstract}

\keywords{X-ray astronomical optics -- Wolter-I optics -- Wide field optics -- Gravitational wave electromagnetic counterparts}

\section{Introduction}

In the last decade gravitational wave detectors have opened a new window on the Universe\cite{PhysRevLett.116.061102, PhysRevLett.119.161101}. To have a complete view of gravitational wave events, we need to also localize and study the electromagnetic counterpart of these events.  
In the next decades new gravitational wave detectors are expected to come into action, observing many more gravitational wave events than is currently possible.
From the future 3rd-generation Einstein Telescope\cite{2023JCAP...07..068B} and Cosmic Explorer\cite{2021arXiv210909882E} 100-300 events/yr are expected, of these $\sim$1\% are expected to have a localization of $\sim$10~deg$^2$, and $\sim$10\% are expected to have a localization of $\sim$100~deg$^2$. With the space-based Laser Interferometer Space Antenna (LISA\cite{2017arXiv170200786A}), the predictions are for error regions of the order of 10--100~deg$^2$.

To look for the counterpart in X-rays quickly enough to detect, for example, the cocoon afterglow of the associated off-axis gamma-ray burst, we would need a field of view big enough to be able to observe, with few repointings, the entire region of the sky constrained by gravitational wave detectors. 

X-ray imaging for astronomical applications has been possible, in the last half-century, using the Wolter-I grazing-incidence telescopes. Due to their high energy, X-rays are reflected only at small angles, but optics with single reflection at small angles exhibit strong aberrations: two reflections are necessary to create good quality images. In the Wolter-I design, X-rays are reflected first on a parabolic and then on a hyperbolic surface. With this design, however, just a few tens of arcmin off-axis, the effective area decreases and strong aberrations are introduced. Most X-ray missions in the last decades have combined Wolter-I X-ray optics modules with focal plane instruments --- such as imagers, spectrographs and recently polarimeters --- to study the properties of X-ray sources in a narrow field of view (of the order of hundreds arcmin$^ {2}$), and with good angular resolution (up to a fraction of arcsec). The best performance achieved to date for Wolter-I optics in orbit is a $\sim$3~deg$^2$ field of view and $\sim$ 150 arcsec half energy width at 1~deg off-axis radius with \emph{ROSAT}.

Alternatively some current X-ray telescopes, using other optical designs, such as Lobsters, have a field of view large enough to survey the entire sky, but somewhat worse angular resolution and sensitivity. Wide field instruments have also been built using coded masks, but the sensitivity in angular resolution and background is bad.

In this paper we describe a variation of the Wolter-I design that increases its field of view without increasing the size and mass of the system. This takes advantage of the fact that, to increase collecting area, commonly Wolter-I systems are made of many confocal shells. Our innovation lies in strategically tilting these individual shells in different directions, so increasing the field of view --- at the cost of a decrease of the on-axis effective area and a worsening of the on-axis angular resolution.

\section{Telescope Requirements for Gravitational Wave Counterpart Accurate Localization \label{sec:localization_requirements}}
Table \ref{tab:mission_reqs} reports some proposed requirements for a mission dedicated to accurately localizing gravitational wave events electromagnetic counterparts. The effective area can be set so that an event such as GW170817 can have its cocoon afterglow detectable. The half energy width is set to make it easy to localize the event in a region small enough that optical ground-based telescopes can promptly know its position accurately enough to find it. The field of view is defined as the radial distance off-axis at which the other requirements are still met, and is set to allow for just a few repointings to explore the entire 10--100~deg$^2$ region expected from present and future gravitational wave detectors.

\begin{table*}
    \centering
    \begin{tabular}{c|c|c}
         &  Minimum& Best\\\hline
         Field of view&  $>$10~deg$^2$ = 1.8~deg radius& \\\hline
         Effective Area&  $>$100~cm$^2$ over the field & $>$200~cm$^2$ over the field  \\\hline
         Half Energy Width&  $<$200~arcsec over the field  & $<$100~arcsec over the field 
    \end{tabular}
    \caption{Possible mission requirements for a mission dedicated to accurately localizing gravitational wave events electromagnetic counterparts.}
    \label{tab:mission_reqs}
\end{table*}

\section{Tilted Wolter-I Design \label{sec:tilted_design_basics}}
A standard Wolter-I shell consists of a circular (rotating all around the optical axis) parabola followed by a circular hyperbola, described with equations 
\begin{equation}
R_{parabola}^2 =  R_0^2  -2 R_0 \tan(\theta) z 
\end{equation}
\begin{equation}
R_{iperbola}^2 =  R_0^2  -2 R_0 \tan(\beta) z + \frac{2R_0\tan(\beta)}{f+R_0\cot(2\theta)}
\end{equation}
where, for highest reflectivity\cite{2021hai4.book....3P}, $f=R_0/\tan(4\theta)$ and $\beta=3\theta$.

\begin{figure}
    \centering
    \includegraphics[width=0.3\linewidth]{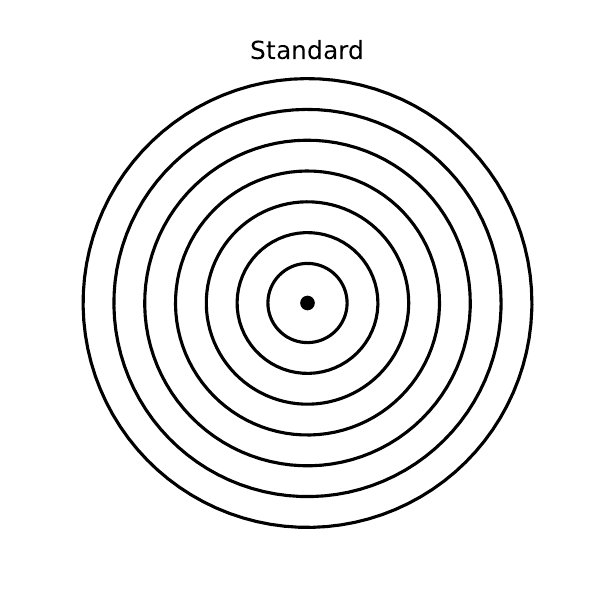}
    \includegraphics[width=0.3\linewidth]{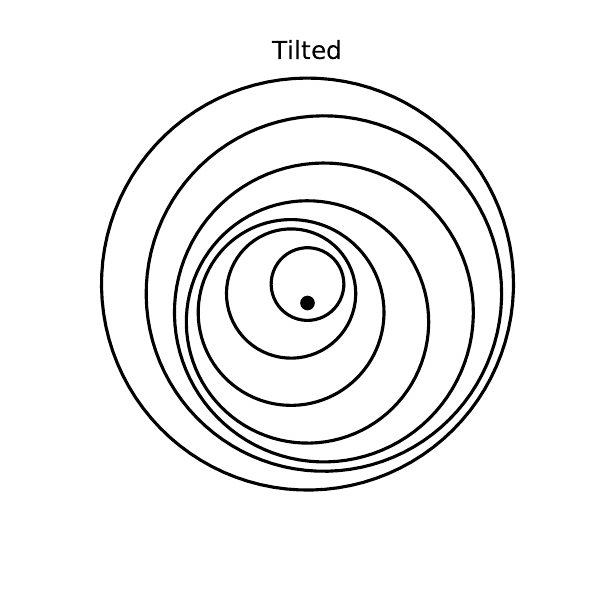}
    \caption{Exemplary standard (left) and tilted (right) Wolter-I shells seen from above.}
    \label{fig:drawings}
\end{figure}

Commonly, Wolter-I systems are made of many confocal shells. This configuration gives the maximum effective area, and the best angular resolution, on-axis, while the performance decreases rapidly off-axis. 
The effective area is the sum of all the effective areas of the single shells so, if all shells are aligned, it is maximized on-axis. If the different shells were to point in different directions, instead, the effective area on-axis will be smaller, due to the smaller contributions by the relative off-axes of the shells, but the field of view will be larger. This is the innovation we study in this work. This can be seen in figure \ref{fig:drawings}.

We simulate the optical configuration and performances in this paper using geometrical optics ray tracing, which describes rays as lines, surfaces by their geometrical equations, and the interaction of X-rays with materials using the Fresnel and Snell's equations. In the design considered in this paper we simulated a reflective coating made of gold with on top a 80~nm layer of carbon --- enhancing reflectivity at low energies. 
We then compute some benchmarks to quantify the performance: the effective area --- computed as the product of the geometrical input area and the fraction of photons arriving on the detector after two reflections (i.e. telescope efficiency) --- and the half energy width --- computed as the diameter of the circle containing half of the impinging photons --- which encapsulates the information on the angular resolution. In the computation of the effective area we assumed a 10~\% absorption due to the structure of the mounting spiders.

In this work we take as starting point the shells used in each of the eROSITA mission optical modules\cite{2021A&A...647A...1P}, but with a shorter length while keeping the same radii --- so allowing to re-use the same mandrels in an hypothetical production phase. Table \ref{tab:mission_proposal} shows the parameters we used in the simulations in this work.

We describe a rotation using the tilt angle $\rho$, which indicates the tilt of the shell from the optical axis, and the position angle $\phi$, which indicates the rotation around the optical axis. 
To simulate the effect of tilting the shells in practice, we use the ray-tracing software to perform an equivalent transformation on the incoming rays. For a shell tilted by an angle $(\rho, \phi)$, an incoming ray is first rotated by $(-\rho, -\phi)$ relative to the top of the optics. After the ray interacts with the optics, it is rotated back by $(\rho, \phi)$ relative to the bottom of the optics.

The shells not only are tilted by the two angles $\rho$ and $\phi$ defined above, but also need to be shifted in the $x-y$ plane so that their focal points are all centered in the same point (example values are reported in the field ``Shells relative horizontal shifts'' in table \ref{tab:mission_proposal}).

\begin{table}
    \centering
    \begin{tabular}{c|c}
         Number of shells& 54\\\hline
         Diameter min& 75~mm\\\hline
         Diameter max& 348~mm\\\hline
         Mirror lengths& 90~mm+90~mm\\\hline
         Focal distance& 1600~mm\\\hline
         Best focus distance& 1595.4~mm\\\hline
         Tilt angle& 0.9$^\circ$\\\hline
 	 Shells relative horizontal shifts&0-0.5~mm\\\hline
         Weight& 10~kg
    \end{tabular}
    \caption{Parameters of a possible module, based on tilted Wolter-I shells, for a fast-class mission based on the requirements of table \ref{tab:mission_reqs} and under 20~kg weight (so that two modules can be used). The shells of each of the eROSITA mission optical modules\cite{2021A&A...647A...1P} have been used for the radii, so that the same mandrels could be re-used even if with shorter length compared to eROSITA.}
    \label{tab:mission_proposal}
\end{table}

The simulations performed in this paper were done at 1~keV. At higher energies the critical reflection angle becomes less gracing, and the reflection efficiency decreases. This is particularly true for the tilted design in which, due to the tilt, the incidence angle can be further from grazing than for standard Wolter-I systems. A future work will focus on using appropriate multilayer coatings to extend the energy band higher.

\section{Optimizing the Tilted Design \label{sec:tilted_design_optimization}}

\subsection{Variations of Tilt\label{subsec:tilt}}
We start by comparing different tilt angles with the position angle $\phi$ sampled from multiple cycles from 0 to 360$^\circ$ --- ensuring that the radii of the individual shells are uniformly distributed in the different tilt directions. We also tested a design in which the tilt angle $\rho$ precesses with the position angle $\phi$, but this didn't improve the performance. 

\begin{figure}
    \centering
    \includegraphics[width=0.7\linewidth]{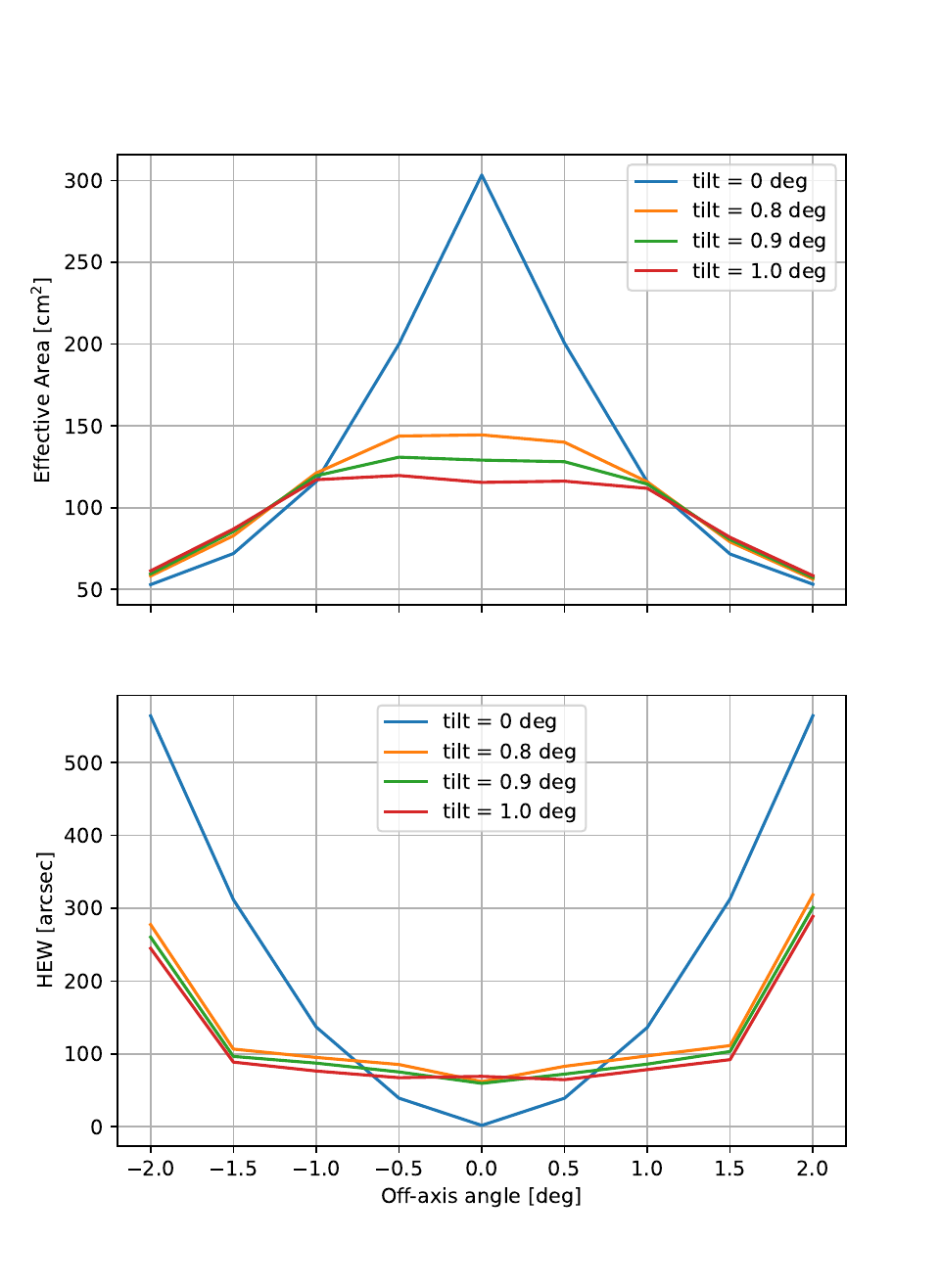}
    \caption{Effective area and half energy width for different tilts at energy 1~keV and best focusing distance.
    The other optical parameters are as reported in table \ref{tab:mission_proposal}.
    }
    \label{fig:tilts}
\end{figure}

Figure \ref{fig:tilts} shows the effective area and half energy width for different tilts: we see that, varying the tilt, the effective area in the center strongly increases, but the angular resolution and effective area off-axis also worsen; based on these simulations we choose, for the remainder of this paper, a tilt angle of $\rho=0.9^\circ$. 

\subsection{Variations of Mirror Length}

\begin{figure}
    \centering
    \includegraphics[width=.7\linewidth]{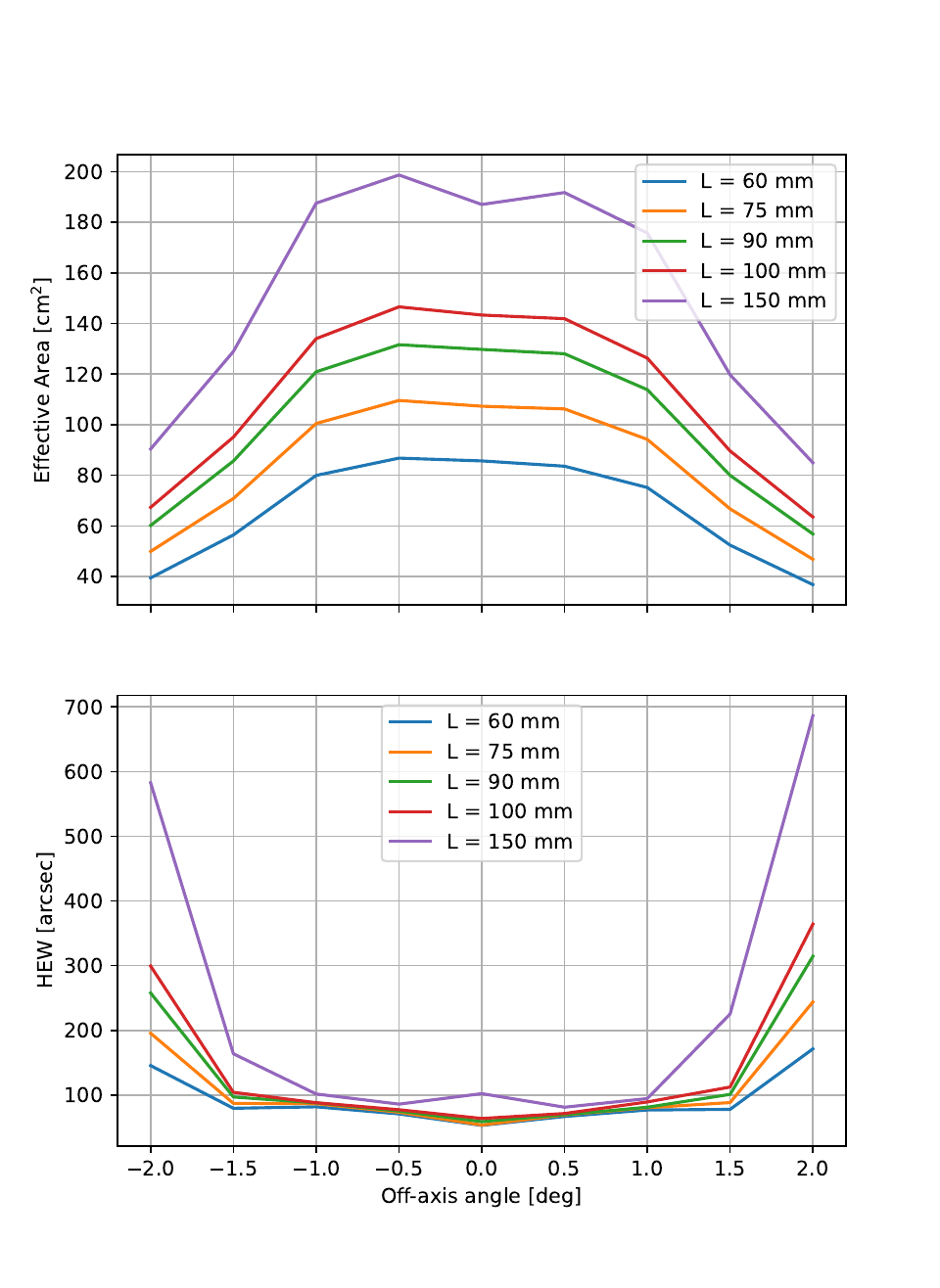}
    \caption{Effective area and half energy width for different mirror lengths of each parabola or hyperbola (segment, so the total length of the mirror module is $2L$) at energy 1~keV and best focusing distance. The other optical parameters are as reported in table \ref{tab:mission_proposal}.}
    \label{fig:tilting_different_L}
\end{figure}

Figure \ref{fig:tilting_different_L} shows the performance with different mirror lengths: longer mirrors increase the effective area, but also worsen the angular resolution. Longer mirrors also cause a significantly heavier mass. 
For the remainder of this paper we choose $L=90$~mm. 

\subsection{Best Focusing Distance}
To transform a standard Wolter-I telescope into a telescope optimized for wide fields, another transformation we need to do is to optimize the focusing distance of the detector, so to worsen the angular resolution on-axis but improve it off-axis: to find the best overall value, we define the figure of merit
\begin{equation}
    FOM = \sum fov^2 \cdot hew \label{eq:FOM_focusing}
\end{equation}
where $fov$ is the field of view (i.e. the off-axis angle) of each element in the sum and $hew$ the half energy width. For the parameters of table \ref{tab:mission_proposal}, a best focus value of 1595.4~mm is found.

\subsection{Ghost Rays and Background}
In this design, compared to standard Wolter-I systems, there is a big fraction of rays that do not focus, but contribute to the background (stray light/unreflected photons) or form ``ghost rays'' (single reflections).

\begin{figure}
    \centering
    \includegraphics[width=0.49\linewidth]{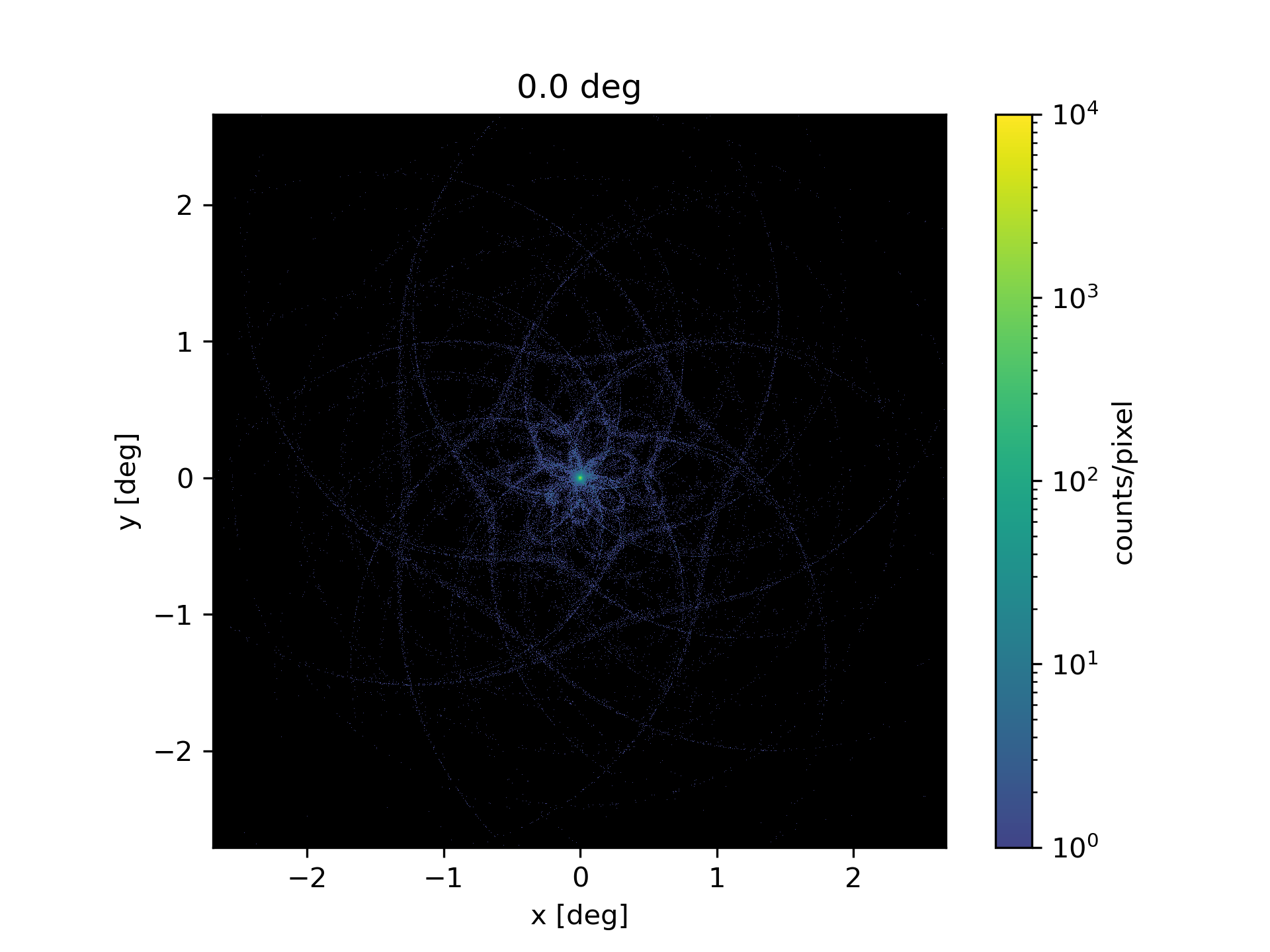}
    \includegraphics[width=0.49\linewidth]{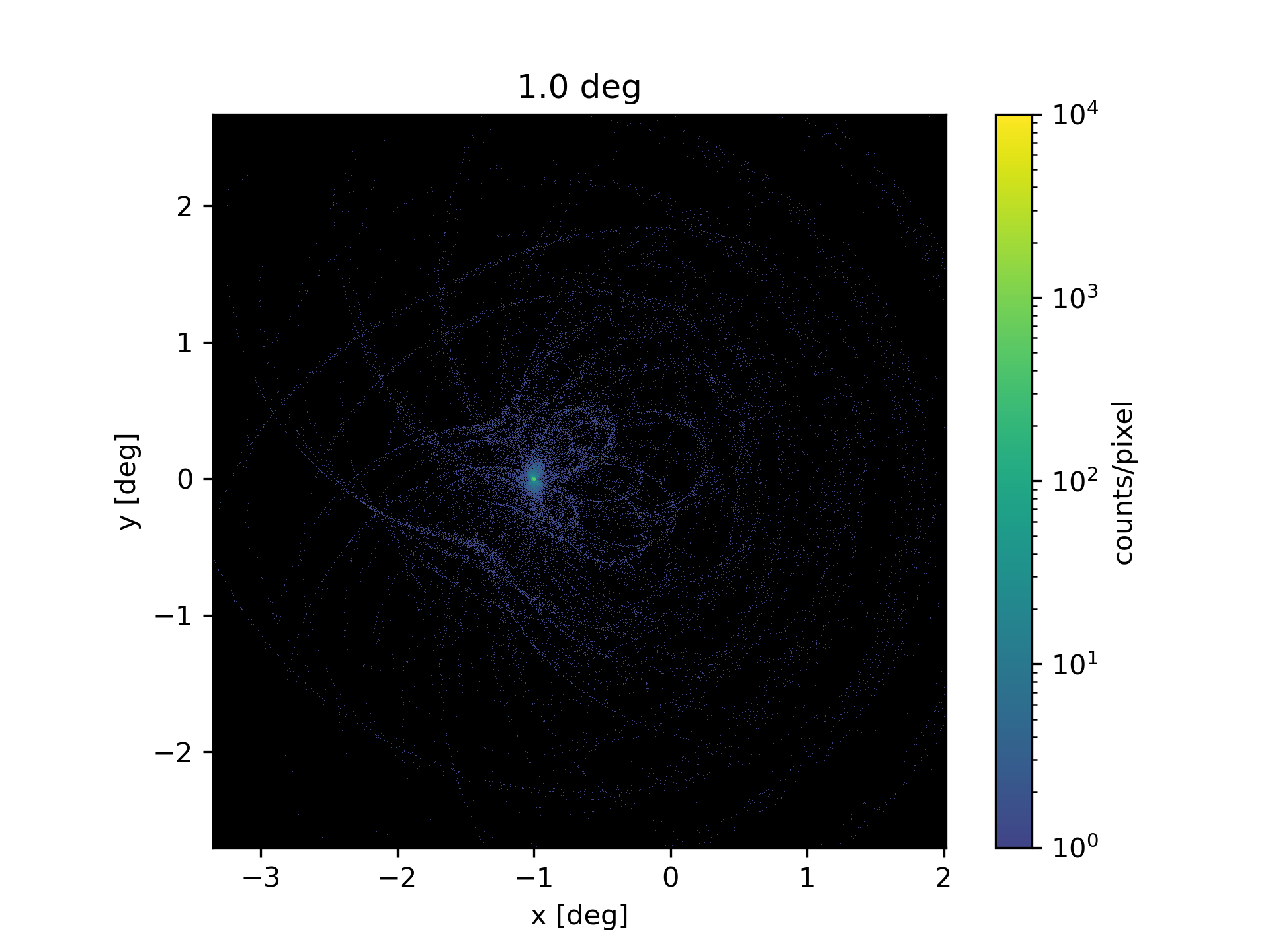}
    \caption{Photon distribution of the configuration considered in this paper, on-axis and off-axis, showing the distribution of ghost rays.} 
\label{fig:PSFs}
\end{figure}

The resulting shape of the photon distribution is shown in figure \ref{fig:PSFs}: the `flower'' shape is produced by the different tiltings of the optics, due to the shape of the point spread function obtained by the combination of shells tilted in different directions. We see that the distribution is over a wide area, but there are still some parts of the ghost rays that could cause confusion with real --- fainter --- sources, in particular where the curved contributions from different shells intersect in almost point-like shapes. The consequences, and ways of mitigating this effect (i.e. baffles), will be analyzed in more detail in future works.

\section{A Fast Mission based on the Tilted Wolter-I Design \label{sec:tilted_mission_proposal}}

For a fast-class astrophysical mission that satisfies the mission requirements outlined in table \ref{tab:mission_reqs}, in particular with a field of view of at least 10~deg$^2$, the main constraining factor is the weight allocated to the optics. This is a value that can vary strongly depending on the total weight allocatable to the satellite payload and also on how the payload is built. In this design, we arbitrarily assume that we can allocate 20~kg to the optics. We assumed that the shells will be manufactured using nickel electroforming replication, a ready-off-the-shelf (TRL=9) technology developed in Italy that has been successfully employed in making the mirrors of several telescopes, including \textit{BeppoSAX}, \textit{XMM-Newton}, \textit{Swift}, eROSITA, and \textit{ Einstein} \textit{Probe}/FXT. The production of tilted Wolter-I shells requires only a slight modification of the standard Wolter-I design; this also makes it possible to reuse mandrels from other missions.

The mirror thickness depends on the length of the mirrors and the desired on-axis half-energy width. Thus, given our relaxed requirements regarding the on-axis HEW, the shells can be made significantly thinner compared to previous missions. Compared to eROSITA (thickness/radius of 0.005\cite{2021A&A...647A...1P}), we can assume a ratio of 0.00175, with a very relevant gain in terms of mass saving.

\begin{figure}
    \centering
    \includegraphics[width=.7\linewidth]{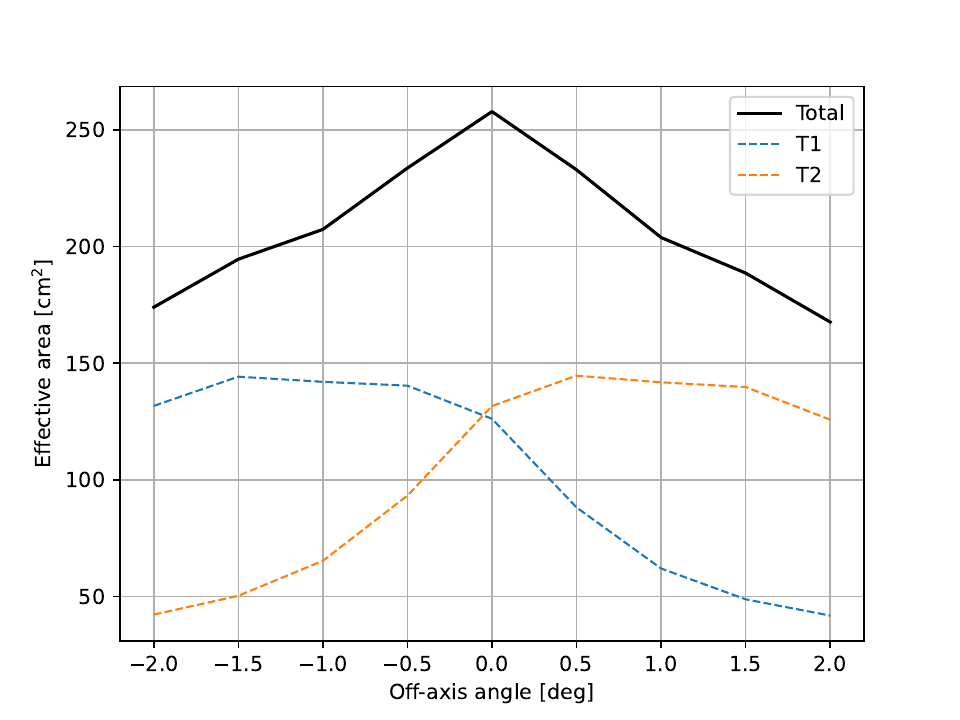}
    \includegraphics[width=.7\linewidth]{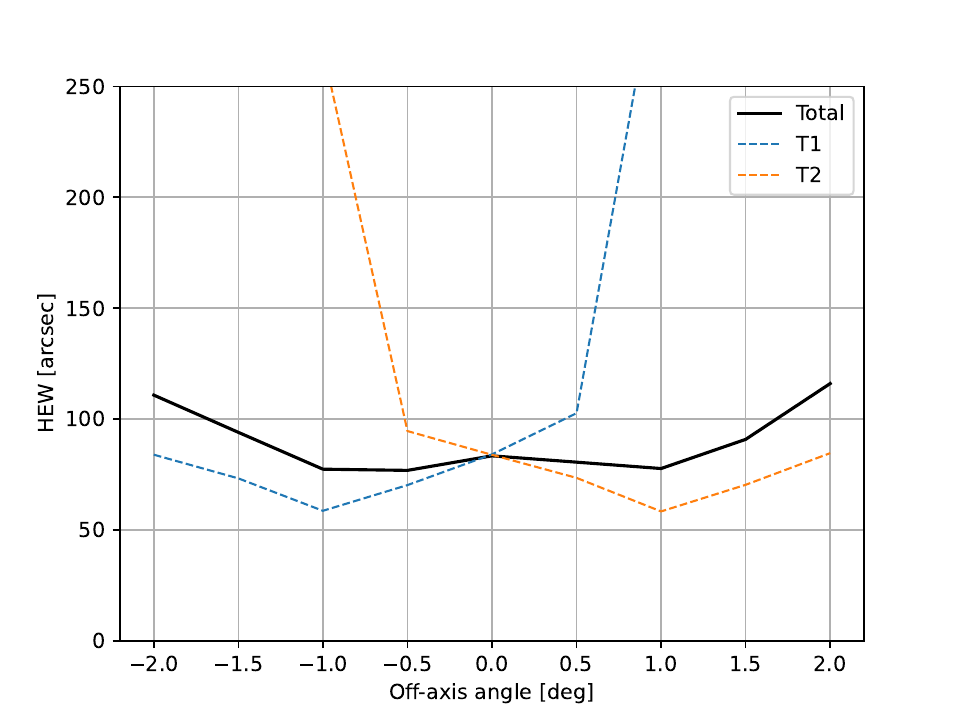}
    \caption{Effective area and half energy width by combining two of the modules considered in this paper (the optical parameters for each are as reported in table \ref{tab:mission_proposal}) tilted 1~deg in opposite directions.}
    \label{fig:two_modules}
\end{figure}

Considering these limits and constraints, we have drawn a mission design which we call WEDGE (Wide-field Explorer for Discovering Gravitational wave Electromagnetic counterparts), already proposed to the Italian Space Agency as small astrophysical mission. The baseline design of a WEDGE module consists of 54 shells produced using the same mandrels as eROSITA, whose details are outlined in table \ref{tab:mission_proposal}. Because the single module has a weight of 10~kg, but we can allocate double the mass, WEDGE uses two of these modules: the combination of two modules tilted by 1~deg in opposite directions is shown in figure \ref{fig:two_modules}.

\section{Conclusion}
In this work, we have shown how introducing a tilt to standard Wolter-I shells expands the field of view, making the optical design suitable for the rapid localization of gravitational wave electromagnetic counterparts. We proposed the WEDGE mission concept, which achieves the required wide-field performance while using using eROSITA mandrels and high-TRL electroforming techniques. Future studies will focus on the mitigation of stray light and ghost rays.

\acknowledgments 
 
This work was supported by fundamental research financing by the Italian National Institute of Astrophysics (INAF): MG-RSN5 1.05.24.07.05

\bibliography{tilted_citations} 

@ARTICLE{2023JCAP...07..068B,
       author = {{Branchesi}, Marica and {Maggiore}, Michele and {Alonso}, David and {Badger}, Charles and {Banerjee}, Biswajit and {Beirnaert}, Freija and {Belgacem}, Enis and {Bhagwat}, Swetha and {Boileau}, Guillaume and {Borhanian}, Ssohrab and {Brown}, Daniel David and {Leong Chan}, Man and {Cusin}, Giulia and {Danilishin}, Stefan L. and {Degallaix}, Jerome and {De Luca}, Valerio and {Dhani}, Arnab and {Dietrich}, Tim and {Dupletsa}, Ulyana and {Foffa}, Stefano and {Franciolini}, Gabriele and {Freise}, Andreas and {Gemme}, Gianluca and {Goncharov}, Boris and {Ghosh}, Archisman and {Gulminelli}, Francesca and {Gupta}, Ish and {Kumar Gupta}, Pawan and {Harms}, Jan and {Hazra}, Nandini and {Hild}, Stefan and {Hinderer}, Tanja and {Siong Heng}, Ik and {Iacovelli}, Francesco and {Janquart}, Justin and {Janssens}, Kamiel and {Jenkins}, Alexander C. and {Kalaghatgi}, Chinmay and {Koroveshi}, Xhesika and {Li}, Tjonnie G.~F. and {Li}, Yufeng and {Loffredo}, Eleonora and {Maggio}, Elisa and {Mancarella}, Michele and {Mapelli}, Michela and {Martinovic}, Katarina and {Maselli}, Andrea and {Meyers}, Patrick and {Miller}, Andrew L. and {Mondal}, Chiranjib and {Muttoni}, Niccol{\`o} and {Narola}, Harsh and {Oertel}, Micaela and {Oganesyan}, Gor and {Pacilio}, Costantino and {Palomba}, Cristiano and {Pani}, Paolo and {Pasqualetti}, Antonio and {Perego}, Albino and {P{\'e}rigois}, Carole and {Pieroni}, Mauro and {Piccinni}, Ornella Juliana and {Puecher}, Anna and {Puppo}, Paola and {Ricciardone}, Angelo and {Riotto}, Antonio and {Ronchini}, Samuele and {Sakellariadou}, Mairi and {Samajdar}, Anuradha and {Santoliquido}, Filippo and {Sathyaprakash}, B.~S. and {Steinlechner}, Jessica and {Steinlechner}, Sebastian and {Utina}, Andrei and {Van Den Broeck}, Chris and {Zhang}, Teng},
        title = "{Science with the Einstein Telescope: a comparison of different designs}",
      journal = {\jcap},
         year = 2023,
        month = jul,
       volume = {2023},
       number = {7},
          eid = {068},
        pages = {068},
          doi = {10.1088/1475-7516/2023/07/068},
archivePrefix = {arXiv},
       eprint = {2303.15923},
 primaryClass = {gr-qc},
       adsurl = {https://ui.adsabs.harvard.edu/abs/2023JCAP...07..068B}
}

@INCOLLECTION{2021hai4.book....3P,
       author = {{Pareschi}, Giovanni and {Spiga}, Daniele and {Pelliciari}, Carlo},
        title = "{X-ray Telescopes Based on Wolter-I Optics}",
    booktitle = {The WSPC Handbook of Astronomical Instrumentation, Volume 4: X-Ray Astronomical Instrumentation,},
         year = 2021,
        pages = {3-31},
          doi = {10.1142/9789811203800_0001},
       adsurl = {https://ui.adsabs.harvard.edu/abs/2021hai4.book....3P}
}

@ARTICLE{2021A&A...647A...1P,
       author = {{Predehl}, P. and {Andritschke}, R. and {Arefiev}, V. and {Babyshkin}, V. and {Batanov}, O. and {Becker}, W. and {B{\"o}hringer}, H. and {Bogomolov}, A. and {Boller}, T. and {Borm}, K. and {Bornemann}, W. and {Br{\"a}uninger}, H. and {Br{\"u}ggen}, M. and {Brunner}, H. and {Brusa}, M. and {Bulbul}, E. and {Buntov}, M. and {Burwitz}, V. and {Burkert}, W. and {Clerc}, N. and {Churazov}, E. and {Coutinho}, D. and {Dauser}, T. and {Dennerl}, K. and {Doroshenko}, V. and {Eder}, J. and {Emberger}, V. and {Eraerds}, T. and {Finoguenov}, A. and {Freyberg}, M. and {Friedrich}, P. and {Friedrich}, S. and {F{\"u}rmetz}, M. and {Georgakakis}, A. and {Gilfanov}, M. and {Granato}, S. and {Grossberger}, C. and {Gueguen}, A. and {Gureev}, P. and {Haberl}, F. and {H{\"a}lker}, O. and {Hartner}, G. and {Hasinger}, G. and {Huber}, H. and {Ji}, L. and {Kienlin}, A. v. and {Kink}, W. and {Korotkov}, F. and {Kreykenbohm}, I. and {Lamer}, G. and {Lomakin}, I. and {Lapshov}, I. and {Liu}, T. and {Maitra}, C. and {Meidinger}, N. and {Menz}, B. and {Merloni}, A. and {Mernik}, T. and {Mican}, B. and {Mohr}, J. and {M{\"u}ller}, S. and {Nandra}, K. and {Nazarov}, V. and {Pacaud}, F. and {Pavlinsky}, M. and {Perinati}, E. and {Pfeffermann}, E. and {Pietschner}, D. and {Ramos-Ceja}, M.~E. and {Rau}, A. and {Reiffers}, J. and {Reiprich}, T.~H. and {Robrade}, J. and {Salvato}, M. and {Sanders}, J. and {Santangelo}, A. and {Sasaki}, M. and {Scheuerle}, H. and {Schmid}, C. and {Schmitt}, J. and {Schwope}, A. and {Shirshakov}, A. and {Steinmetz}, M. and {Stewart}, I. and {Str{\"u}der}, L. and {Sunyaev}, R. and {Tenzer}, C. and {Tiedemann}, L. and {Tr{\"u}mper}, J. and {Voron}, V. and {Weber}, P. and {Wilms}, J. and {Yaroshenko}, V.},
        title = "{The eROSITA X-ray telescope on SRG}",
      journal = {\aap},
         year = 2021,
        month = mar,
       volume = {647},
          eid = {A1},
        pages = {A1},
          doi = {10.1051/0004-6361/202039313},
archivePrefix = {arXiv},
       eprint = {2010.03477},
 primaryClass = {astro-ph.HE},
       adsurl = {https://ui.adsabs.harvard.edu/abs/2021A&A...647A...1P}
}

@ARTICLE{2017arXiv170200786A,
       author = {{Amaro-Seoane}, Pau and {Audley}, Heather and {Babak}, Stanislav and {Baker}, John and {Barausse}, Enrico and {Bender}, Peter and {Berti}, Emanuele and {Binetruy}, Pierre and {Born}, Michael and {Bortoluzzi}, Daniele and {Camp}, Jordan and {Caprini}, Chiara and {Cardoso}, Vitor and {Colpi}, Monica and {Conklin}, John and {Cornish}, Neil and {Cutler}, Curt and {Danzmann}, Karsten and {Dolesi}, Rita and {Ferraioli}, Luigi and {Ferroni}, Valerio and {Fitzsimons}, Ewan and {Gair}, Jonathan and {Gesa Bote}, Lluis and {Giardini}, Domenico and {Gibert}, Ferran and {Grimani}, Catia and {Halloin}, Hubert and {Heinzel}, Gerhard and {Hertog}, Thomas and {Hewitson}, Martin and {Holley-Bockelmann}, Kelly and {Hollington}, Daniel and {Hueller}, Mauro and {Inchauspe}, Henri and {Jetzer}, Philippe and {Karnesis}, Nikos and {Killow}, Christian and {Klein}, Antoine and {Klipstein}, Bill and {Korsakova}, Natalia and {Larson}, Shane L and {Livas}, Jeffrey and {Lloro}, Ivan and {Man}, Nary and {Mance}, Davor and {Martino}, Joseph and {Mateos}, Ignacio and {McKenzie}, Kirk and {McWilliams}, Sean T and {Miller}, Cole and {Mueller}, Guido and {Nardini}, Germano and {Nelemans}, Gijs and {Nofrarias}, Miquel and {Petiteau}, Antoine and {Pivato}, Paolo and {Plagnol}, Eric and {Porter}, Ed and {Reiche}, Jens and {Robertson}, David and {Robertson}, Norna and {Rossi}, Elena and {Russano}, Giuliana and {Schutz}, Bernard and {Sesana}, Alberto and {Shoemaker}, David and {Slutsky}, Jacob and {Sopuerta}, Carlos F. and {Sumner}, Tim and {Tamanini}, Nicola and {Thorpe}, Ira and {Troebs}, Michael and {Vallisneri}, Michele and {Vecchio}, Alberto and {Vetrugno}, Daniele and {Vitale}, Stefano and {Volonteri}, Marta and {Wanner}, Gudrun and {Ward}, Harry and {Wass}, Peter and {Weber}, William and {Ziemer}, John and {Zweifel}, Peter},
        title = "{Laser Interferometer Space Antenna}",
      journal = {arXiv e-prints},
         year = 2017,
        month = feb,
          eid = {arXiv:1702.00786},
        pages = {arXiv:1702.00786},
          doi = {10.48550/arXiv.1702.00786},
archivePrefix = {arXiv},
       eprint = {1702.00786},
 primaryClass = {astro-ph.IM},
       adsurl = {https://ui.adsabs.harvard.edu/abs/2017arXiv170200786A}
}

@article{PhysRevLett.116.061102,
  title = {Observation of Gravitational Waves from a Binary Black Hole Merger},
author = {{Abbott B. P. et al. (LIGO Scientific Collaboration and Virgo Collaboration)}},  note = {LIGO Scientific Collaboration and Virgo Collaboration},
  journal = {Phys. Rev. Lett.},
  volume = {116},
  issue = {6},
  pages = {061102},
  numpages = {16},
  year = {2016},
  month = {Feb},
  publisher = {American Physical Society},
  doi = {10.1103/PhysRevLett.116.061102},
  url = {https://link.aps.org/doi/10.1103/PhysRevLett.116.061102}
}

@article{PhysRevLett.119.161101,
  title = {GW170817: Observation of Gravitational Waves from a Binary Neutron Star Inspiral},
author = {{Abbott B. P. et al. (LIGO Scientific Collaboration and Virgo Collaboration)}},  note = {LIGO Scientific Collaboration and Virgo Collaboration},
  journal = {Phys. Rev. Lett.},
  volume = {119},
  issue = {16},
  pages = {161101},
  numpages = {18},
  year = {2017},
  month = {Oct},
  publisher = {American Physical Society},
  doi = {10.1103/PhysRevLett.119.161101},
  url = {https://link.aps.org/doi/10.1103/PhysRevLett.119.161101}
}

@ARTICLE{2021arXiv210909882E,
       author = {{Evans}, Matthew and {Adhikari}, Rana X and {Afle}, Chaitanya and {Ballmer}, Stefan W. and {Biscoveanu}, Sylvia and {Borhanian}, Ssohrab and {Brown}, Duncan A. and {Chen}, Yanbei and {Eisenstein}, Robert and {Gruson}, Alexandra and {Gupta}, Anuradha and {Hall}, Evan D. and {Huxford}, Rachael and {Kamai}, Brittany and {Kashyap}, Rahul and {Kissel}, Jeff S. and {Kuns}, Kevin and {Landry}, Philippe and {Lenon}, Amber and {Lovelace}, Geoffrey and {McCuller}, Lee and {Ng}, Ken K.~Y. and {Nitz}, Alexander H. and {Read}, Jocelyn and {Sathyaprakash}, B.~S. and {Shoemaker}, David H. and {Slagmolen}, Bram J.~J. and {Smith}, Joshua R. and {Srivastava}, Varun and {Sun}, Ling and {Vitale}, Salvatore and {Weiss}, Rainer},
        title = "{A Horizon Study for Cosmic Explorer: Science, Observatories, and Community}",
      journal = {arXiv e-prints},
         year = 2021,
        month = sep,
          eid = {arXiv:2109.09882},
        pages = {arXiv:2109.09882},
          doi = {10.48550/arXiv.2109.09882},
archivePrefix = {arXiv},
       eprint = {2109.09882},
 primaryClass = {astro-ph.IM},
       adsurl = {https://ui.adsabs.harvard.edu/abs/2021arXiv210909882E}
}
\bibliographystyle{spiebib}

\end{document}